\documentclass[10pt,iop]{emulateapj}
\pdfoutput=1 

\usepackage{graphicx}
\usepackage{float}
\usepackage{amsmath}
\usepackage{epsfig}
\usepackage{color}

\newcommand{\ve}[1]{\mathbf{#1}}

\newcommand{\lm}{{\ell m}}

\newcommand{\D}{\ve{D}}

\newcommand{\lmax}{\ell_\text{max}}

\newcommand{\N}{\ve{N}}

\newcommand{\B}{\ve{B}}
\newcommand{\Y}{\ve{Y}}
\newcommand{\Yobs}{\ve{Y}_\text{obs}}

\newcommand{\W}{\ve{W}}

\renewcommand{\a}{\ve{a}}

\newcommand{\Nside}{N_\text{side}}
\newcommand{\Npix}{N_\text{pix}}
\newcommand{\Nrings}{N_\text{rings}}
\newcommand{\Nbands}{N_\text{bands}}

\begin{document}

\title{SymPix: A spherical grid for efficient sampling of rotationally invariant operators}

\author{D. S. Seljebotn\altaffilmark{1} and H. K. Eriksen\altaffilmark{1}}

\email{d.s.seljebotn@astro.uio.no}

\altaffiltext{1}{Institute of Theoretical Astrophysics, University of
  Oslo, P.O.\ Box 1029 Blindern, N-0315 Oslo, Norway}

\begin{abstract}
  We present SymPix, a special-purpose spherical grid optimized for
  efficient sampling of rotationally invariant linear operators. This
  grid is conceptually similar to the Gauss-Legendre (GL) grid,
  aligning sample points with iso-latitude rings located on Legendre
  polynomial zeros. Unlike the GL grid, however, the number of grid
  points per ring varies as a function of latitude, avoiding expensive
  over-sampling near the poles and ensuring nearly equal sky area per
  grid point. The ratio between the number of grid points in two
  neighbouring rings is required to be a low-order rational number (3,
  2, 1, 4/3, 5/4 or 6/5) to maintain a high degree of symmetries. Our
  main motivation for this grid is to solve linear systems using
  multi-grid methods, and to construct efficient preconditioners
  through pixel-space sampling of the linear operator in question. The
  GL grid is not suitable for these purposes due to its massive
  over-sampling near the poles, leading to nearly degenerate linear
  systems, while HEALPix, another commonly used spherical grid,
  exhibits few symmetries, and is therefore computationally
  inefficient for these purposes. As a benchmark and representative
  example, we compute a preconditioner for a linear system with both
  HEALPix and SymPix that involves the operator $\widehat{\D} +
  \widehat{\B}^{T}\N^{-1}\widehat{\B}$, where $\widehat{\B}$ and
  $\widehat{\D}$ may be described as both local and rotationally
  invariant operators, and $\mathbf{N}$ is diagonal in pixel
  domain. For a bandwidth limit of $\lmax=3000$, we find that SymPix,
  due to its higher number of internal symmetries, yields average
  speed-ups of 360 and 23 for $\widehat{\B}^{T} \N^{-1} \widehat{\B}$
  and $\widehat{\D}$, respectively, relative to HEALPix.
\end{abstract}

\shorttitle{SymPix: Spherical grid for sampling of rotationally invariant operators}

\section{Introduction}
\label{sec:introduction}

Unlike the plane, it is impossible to construct a regular
discretization of the sphere. Instead, every conceivable spherical
grid comes with its own set of trade-offs, emphasizing one or more
features at the cost of others. Thus, there is no such thing as a
perfect spherical grid, but the optimal grid instead depends
sensitively on the application under consideration.

In this paper, we will restrict our attention to high-resolution grids
designed for fast and accurate spherical harmonic transforms
(SHTs). In such cases, the primary consideration is that the grid must
allow for efficient $\mathcal{O}(\lmax^3)$ SHTs, where $\lmax$ denotes the upper
harmonic space bandwidth limit of the field in question, as opposed to
the $\mathcal{O}(\lmax^4)$ scaling resulting from naive brute-force summation.
This requires the use of Fast Fourier Transforms (FFTs) in the
longitudinal direction, which in turn implies that i) sample points
must be placed on a set of iso-latitude rings, and ii) sample points
within each ring must be equidistant. However, there is still
flexibility in choosing the latitude of each ring ($\theta_j
\in [0, \pi]$), the number of grid points along each ring ($n_j$), and
the initial offset of each ring ($\phi_{0,j}$).

Three popular spherical grids are the equiangular grid, the
Gauss-Legendre grid \citep[e.g.,][]{doroshkevich:2005}, and
HEALPix\footnote{http://healpix.sourceforge.org} \citep{healpix}.  Of
these, the equiangular grid is the most straightforward, simply
defined by evenly spaced grid points $(\theta_i, \phi_i)$ in both
directions. This grid is typically used for geographical maps, and it
is therefore also called a geographical grid.

Similarly, the standard Gauss-Legendre grid has a constant number of
grid points per ring. However, the ring latitudes $\theta_j$ are
defined such that $P_{N_\text{rings}}(\cos \theta_j)=0$, where $P_n$ is
the Legendre polynomial of degree $n$. This simple modification allows
efficient spherical harmonic analysis to machine precision, and the
grid is thus optimized for spherical harmonics transforms.

Both of these grids suffers from a massive over-sampling of the polar
regions ($\theta$ close to $0$ or $\pi$) compared to the equatorial
region ($\theta \approx \pi/2$), and this renders them sub-optimal,
and sometimes even useless, for certain practical applications. An
important example is the solution of discretized and bandwidth limited
linear systems. If there is a large number of sample points within the
correlation length implied by $\lmax$, the system becomes degenerate
and numerically unstable. Grids with nearly constant pixel areas
perform much better than grids with strongly varying pixel areas for
this type of applications.

One example of such grids is HEALPix, which is short for
``Hierarchical Equal Area and Latitude Pixelization''. This grid has
by construction both constant area pixel area per pixel and grid
points located on iso-latitude, and is as such a good general-purpose
grid. However, this generality comes at a cost in terms of spherical
harmonics precision, as well as a low level of internal pixel
symmetries.

The latter point is particularly important for our applications.
Consider a function of two grid points, $\hat{n}_1$ and $\hat{n}_2$,
that is both localized and rotationally invariant,
\begin{equation}
  \label{eq:f}
  f(\hat{n}_1, \hat{n}_2) = \left\{ \begin{array}{ll}
    f(\hat{n}_1 \cdot \hat{n}_2) & \text{if $\text{arccos}(\hat{n}_1 \cdot \hat{n}_2) < k \Delta$} \\
    0 & \text{otherwise},
\end{array} \right.
\end{equation}
where $\Delta$ denotes the average distance between two neighbouring
grid points. Thus, $f$ is assumed identically zero if the two grid
points are separated by more than $k$ grid units. In our applications,
which employ multi-grid and/or preconditioning methods, we need to
evaluate $f$ for all relevant pairs $(\hat{n}_1, \hat{n}_2)$.
Furthermore, because $f$ typically is computationally expensive, it is
important to minimize the total number of function evaluations, and
large speed-ups can be gained by exploiting symmetries and caching.

For HEALPix, $f$ needs to be evaluated $\mathcal{O}(k^2 N_\text{pix})$
times, because the angular distances between neighbouring grid points
are all different, up to handful of overall symmetries. In contrast,
for the equi-angular and Gauss-Legendre grids only $\mathcal{O}(k^2
\sqrt{N_\text{pix}})$ evaluations are needed. Since the number of grid
points is constant for every ring, we only need to evaluate $f$ for
the first grid point on every ring, accounting for all its neighbours,
after which all function evaluations along the same ring will be given
by symmetry.

In this paper, we construct a novel spherical grid called SymPix that
combines the spherical harmonics transform precision of the
Gauss-Legendre grid with the nearly uniform sample point distances of
HEALPix, while at the same time maintaining a high degree symmetries
within each ring, ensuring that fully sampling $f(\hat{n_1} \cdot
\hat{n_2})$ scales as $\mathcal{O}(k^2 \sqrt{\Npix})$.

\section{The SymPix grid}
\label{sec:design}

\begin{figure*}
  \begin{center}
    \includegraphics[width=\linewidth]{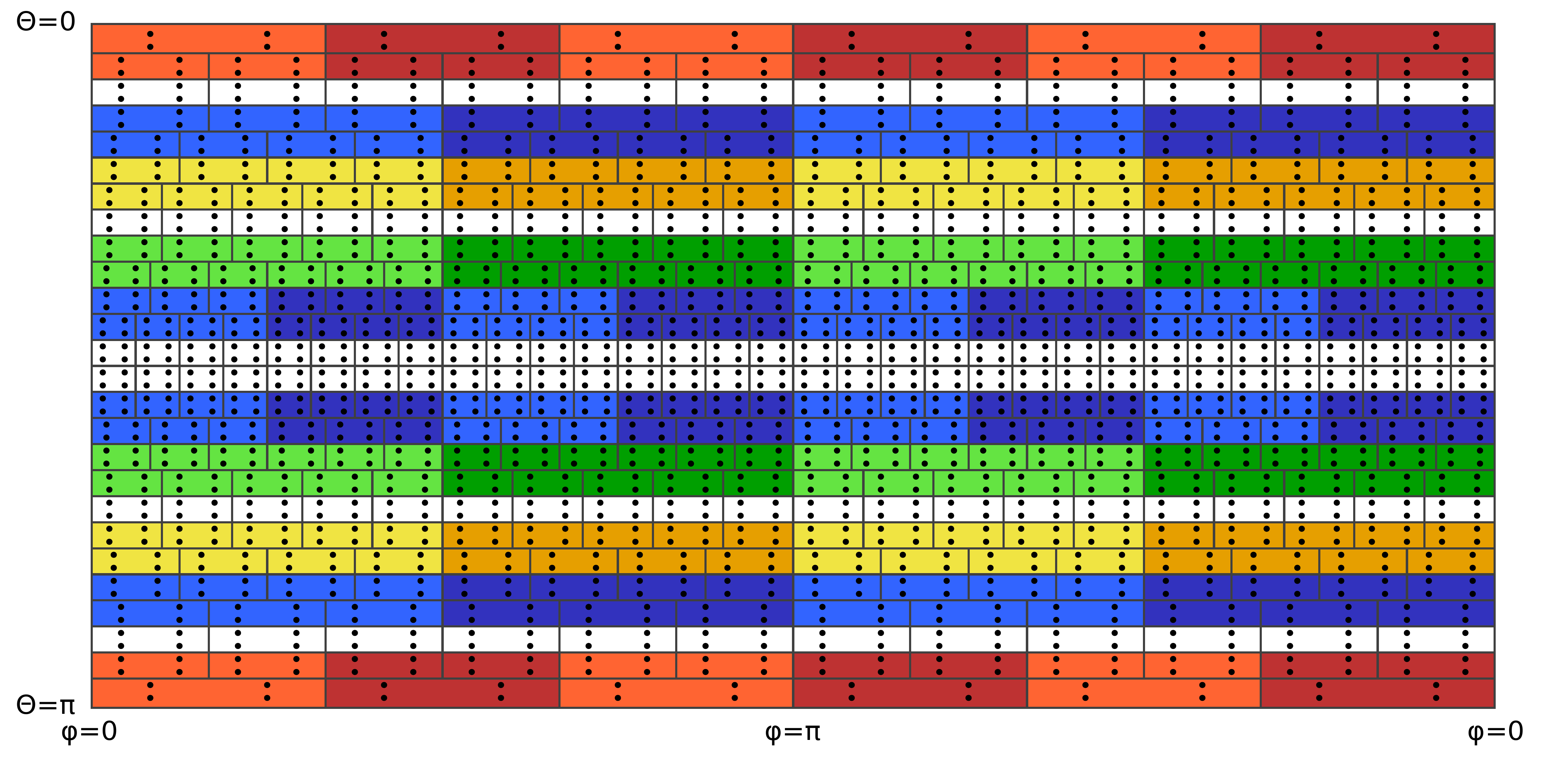}
  \end{center}
  \caption{Geometric layout of SymPix sample points, implementing a
    cylindrical projection of the sphere.  Each rectangle indicates a
    {\em tile} of (in this case) $2 \times 2$ sample points. For white
    tile-bands, the bands above and below have the same number of
    tiles, and angular distances between sample points in a given tile
    and sample points in the neighbouring tiles are therefore constant
    throughout the band. Function evaluations depending only on
    angular distances may therefore be cached and reused.  Colored
    tile-bands increment the number of tiles by a factor of $2$ (red),
    $4/3$ (blue), $5/4$ (yellow), $6/5$ (green), and $4/3$ again
    (blue) towards the equator.  For these bands, the neighbouring
    tile relationship repeats itself (as indicated by shading), and
    there are still only a few cases that need to be computed and
    cached for each band.}
  \label{fig:pattern}
\end{figure*}
\begin{figure*}
  \begin{center}
    \includegraphics[width=\linewidth]{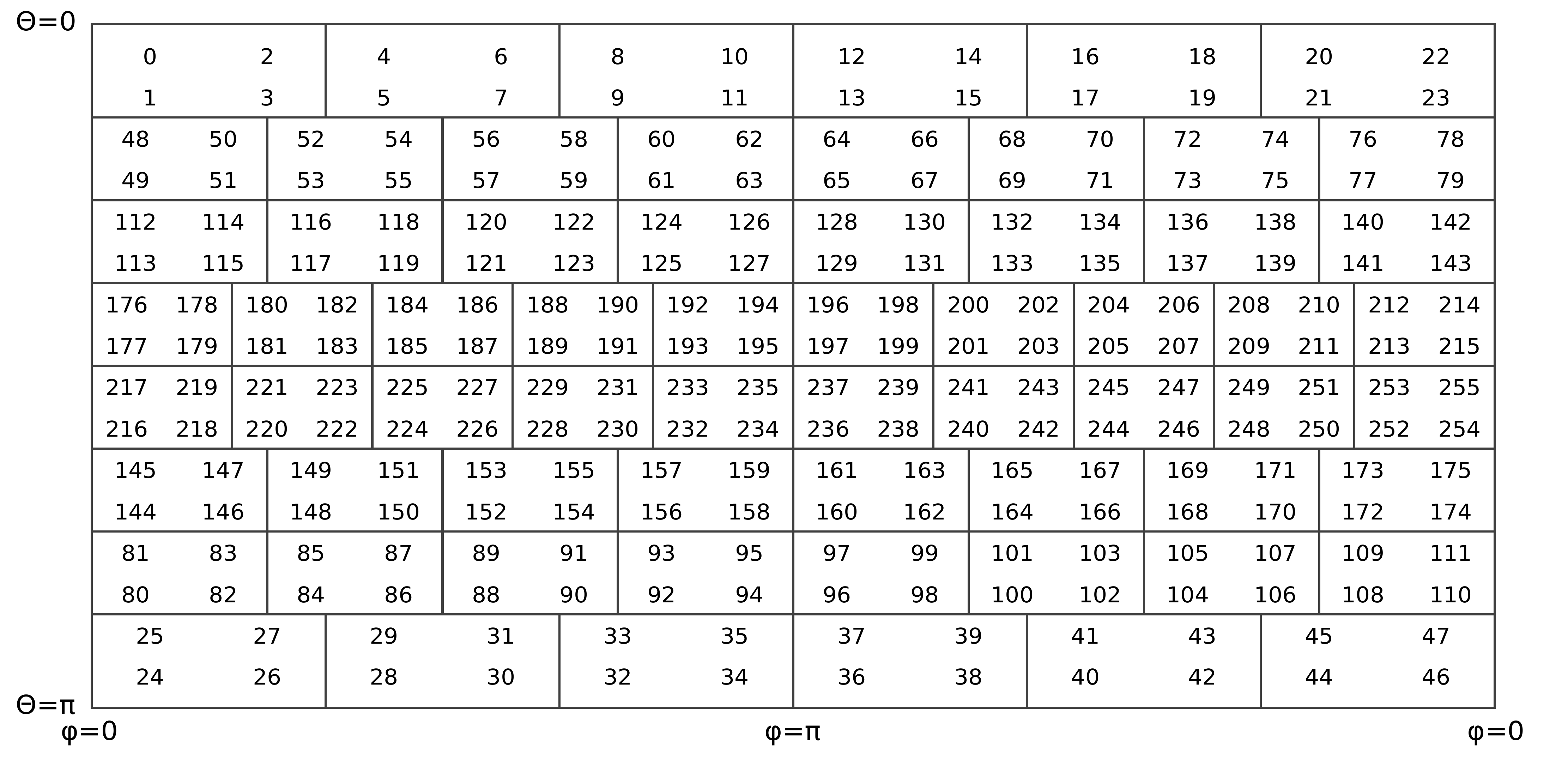}
  \end{center}
  \caption{Memory ordering of SymPix sample points. Note that the
    resolution is lower than in Figure~\ref{fig:pattern}.  Within each
    band the pixel order increases first latitudinally, i.e., along
    the $\theta$ direction. This ensures that access within the same
    tile is local in memory, and there are no discontinuities along
    each ring, which is convenient for SHTs. Additionally, to
    support efficient distributed programming, we interleave Northern
    and Southern bands, such that they naturally are assigned to the
    same node without explicit additional book-keeping.}
  \label{fig:ordering}
\end{figure*}

\subsection{Ring layout basics}
\label{sec:requirements}

The main role of the SymPix grid is that of a supporting grid in
internal multi-grid and/or preconditioning calculations, and
maintaining high numerical precision is therefore essential. For this
reason, we adopt the Gauss-Legendre latitudinal ring layout as the
basis of our grid. This provides support for both spherical harmonic
\emph{synthesis} (i.e., transforming from harmonic coefficients to
pixel space) and \emph{analysis} (transforming from pixel space to
harmonic coefficients) to machine precision, by virtue of having an
exact quadrature rule on the form
\begin{equation}
  \label{eq:shdef}
    a_\lm = \int_{\Omega} Y^*(\hat{n}) f(\hat{n}) d\Omega = \sum_i Y^*(\hat{n}_i) f(\hat{n}_i) w_i,
\end{equation}
where $w_i$ is a set of quadrature weights. By placing rings
exclusively on the zeros of the $\lmax$'th polynomial, one is
guaranteed that $P_{\lmax+1}(\cos \theta_i) = 0$, and the discretized
field is algebraically bandwidth limited to harmonic modes with
$\ell\le\lmax$.

Next, we need to include enough sample points along each ring to fully
resolve all spherical harmonic modes with $\ell\le\lmax$. Formally
speaking, this requires $2 \Nrings$ grid points per ring. However,
this requirement is somewhat counter-intuitive by suggesting massive
over-sampling of the polar regions compared to the equatorial
region. And, indeed, our intuition is correct: The spherical harmonic
modes $Y_\lm(\theta, \phi)$ are very close to zero in the polar
regions for high $\ell$ and $m$, and these are the only modes that can
cause high-frequency variation in the longitudinal direction.  For
this reason, the \verb@libsharp@ SHT package \citep{reinecke:2013}
omits $Y_\lm(\theta, \phi)$ whenever
\begin{equation}
  \label{eq:polar-optimization}
  \sqrt{m^2 - 2 m \cos \theta} - \lmax \sin \theta > \text{max}(100, 0.01\lmax),
\end{equation}
exploiting that contributions from higher-ordered harmonics are
numerically irrelevant. An explicit bound on the number of pixels
required for machine precision was derived by
\citet{prezeau-reinecke-2010}, and \cite{reinecke:2013} used this to
construct the {\em reduced Gauss-Legendre grid}.  Explicitly, for a
given ring located at some latitude $\theta$,
Equation~\ref{eq:polar-optimization} defines the maximum $m$ such
that $Y_\lm(\theta, \phi)$ does not vanish. The minimum number of
pixels on that ring is then given by $2m + 1$, resulting in a
longitudinal sample frequency that exceeds the Nyquist frequency.

\subsection{Tiling}
As discussed in Section~\ref{sec:introduction}, our primary usecase is
evaluating a function $f(\hat{n}_1, \hat{n}_2)$ for all possible pairs
$(\hat{n}_1, \hat{n}_2)$, but with the restriction that $f$ is zero
unless $\hat{n}_1$ and $\hat{n}_2$ are close together. To avoid
unnecessary searches over vanishing pairs, we therefore partition our
grid into a set of $k \times k$-sized tiles, where $k$ is chosen such
that $f(\hat{n}_1, \hat{n}_2) = 0$ unless $\hat{n}_1$ and $\hat{n}_2$
are either in the same tile or in two neighbouring tiles. Thus,
finding all relevant partner points for a given grid point simply
amounts to a closest neighbour tile look-up. However, this also
requires that the number of rings is divisible by $k$ (letting
$\Nrings > \lmax + 1$ if necessary), and that a set of $k$ consequtive
rings must have the same number of sample points.  We will refer to
each such set of $k$ rings as a {\em band}.

\subsection{Enforcing symmetries}
\label{sec:symmetry}

The main remaining step is to define the number of tiles per band. On
the one hand, it must satisfy the minimum number of pixels given by
Equation~\ref{eq:polar-optimization}. On the other hand, it may be
beneficial to increase it beyond this, in order to increase symmetries
within and across bands. For instance, if we sample $f$ from
Equation~\ref{eq:f} for all point-pairs within a tile, the result can
obviously be re-used for all tiles in that band, since all
between-point angular distances are conserved between
tiles. Similarly, we can reuse results between neighbouring tiles
within the same band due to longitudinal symmetry.

In addition, we exploit the additional degrees of freedom in choosing
the number of tiles to ensure symmetries with respect to latitudinally
neighbouring tiles. Specifically, we require that the number of tiles
can increase from one band to the next only by a factor of exactly 3,
2, 1, 4/3, 5/4, or 6/5. Additionally, at least two bands in a row must
have the same number of tiles, except for the polar bands. Finally, in
order to avoid special cases we allow no equatorial ring (i.e., we
insist that $\Nrings$ is an even number), and, purely conventionally,
the location of the first grid point in a given ring is chosen to be
half the pixel distance within that same ring. Together, these
requirements ensure that the pattern of neighbouring tiles repeats
itself with a short period, and the total number of different cases to
evaluate scales as $\mathcal{O}(N_{\textrm{ring}})$ rather than
$\mathcal{O}(N_{\textrm{pix}})$. We employ a dynamic programming
algorithm to find the optimal number of tiles per band, subject to the
constraints defined above, as detailed in Section \ref{sec:algorithm}.
An example grid corresponding to $k=2$ tiling is illustrated in
Figure~\ref{fig:pattern}.

\subsection{Memory layout and pixel ordering}

While the above constraints fully define the geometric properties of
the SymPix grid, they do not imply a canonical memory layout or
``pixel ordering''. To fix this, we adopt two additional rules, both
designed to maximize memory access efficiency and programming
convenience.

First, the Northern and Southern hemispheres are band-wise
interleaved. That is, we first list the Northern-most polar band,
followed by the Southern-most polar band, followed by the second
Northern band and so on. The main advantage of this organization lies
in convenient distributed programming across multiple computing nodes;
interleaving the two hemispheres ensures that the same node can
readily exploit North-South symmetries.

Second, grid points are latitudinally major-ordered within a given
tile, i.e., the pixel ordering increases most rapidly along the
$\theta$ direction. While the order within each tile could have been
in any direction, this choice implies that pixel ordering is
continuous across longitudinal tile borders, which is particularly
convenient for SHTs.

Figure \ref{fig:ordering} provides an example of the resulting pixel
ordering. Note that the resolution is lower than the corresponding
illustration in Figure~\ref{fig:pattern}.

\newcommand{\find}{\quad}
\newcommand{\ind}{\quad \;}
\begin{figure}
  \begin{minipage}{.98\linewidth}

    \begin{tabular}{lcl}
      \multicolumn{3}{l}{Optimal-SymPix-Grid:} \\
      \multicolumn{3}{l}{\find {\bf Inputs:}} \\
      \find  \ind $\lmax$ &--& Band-limit of field to represent \\
      \find  \ind $k$ &--& Tile size \\
      \multicolumn{3}{l}{\find {\bf Output:}} \\
      \find \ind $n$ &--& number of bands \\
      \find \ind $T_i$ &--& number of tiles in each band\\
      \multicolumn{3}{l}{\find {\bf Auxiliary:}} \\
      \find \ind $\alpha_i$ &--& minimum number of tiles for band $i$ \\
      \find \ind $C_{i,t}$ &--& the cost of the best partial solution for \\
      && bands $0$ to $i$ when assuming $T_i=t$ \\
      \find \ind $P_{i,t}$ &--& ``previous-pointers''; when assuming $T_i=t$, \\
      && the solution for bands $0$ to $i$ has $T_{i-1}=P_{i,t}$ \\
      \\
    \end{tabular}
    \begin{tabular}{l}
      \find Treat unassigned $C_{i,t}$ as $\infty$ and unassigned $P_{i,T_0}$ as $-1$ \vspace{0.5mm}\\
      \find $\Nrings$ $\leftarrow$ $\lmax + 1$ rounded up to next
          multiple of $2k$ \\
      \find $n \leftarrow \Nrings / 2k$ \\
      \find Find $\theta_j$ for each ring $j$ as for Gauss-Legendre grid \\
      \find $T \leftarrow \text{max}(100, \lmax/100)$ \\
      \find {\bf for each} $i \in \{0, \dots, n - 1\}$: \\
      \find \ind Find minimum $m$ that satisfies Equation~\eqref{eq:polar-optimization}
                 for $\theta_{ik}$ \\
      \find \ind $\alpha_i \leftarrow \lceil (2m + 1)/k \rceil$ \\
      \find $T_0 \leftarrow \text{min} (\{ 2^i3^j5^k \; | 
          2^i3^j5^k \ge \alpha_0, i \in \mathcal{N}, j \in \mathcal{N}, k \in \mathcal{N} \})$  \vspace{0.5mm}\\

      \find $C_{i,T_0} \leftarrow (T_0 - \alpha_0)^2$ \vspace{0.5mm}\\
      \find {\bf for each} $i$ {\bf from} $1$ {\bf to} $n - 1$: \vspace{0.5mm}\\
      \find \ind {\bf for each} $t_\text{prev}$ such that $C_{i-1,t_\text{prev}} < \infty$: \quad \vspace{0.5mm}\\
      \find \ind \ind {\bf for each} $x \in \{ 3, 2, 1, 4/3, 5/4, 6/5 \}$ such that $xt_\text{prev} \in \mathcal{N}$: \vspace{0.5mm}\\
      \find \ind \ind \ind $t \leftarrow xt_\text{prev}$ \vspace{0.5mm}\\
      \find \ind \ind \ind {\bf if} $C_{i-1,t_\text{prev}} + (\alpha_i - t)^2 < C_{i,t}$ \vspace{0.5mm}\\
      \find \ind \ind \ind \ind \ind and $\alpha_i \le t \le 3\alpha_{i}$ \vspace{0.5mm}\\ 
      \find \ind \ind \ind \ind \ind and ($P_{i-1,t_\text{prev}} = t_\text{prev}$ or $x=1$): \vspace{0.5mm}\\ 
     \find \ind \ind \ind \ind $C_{i,t} \leftarrow c + (\alpha_i - t)^2$  \vspace{0.5mm}\\
      \find \ind \ind \ind \ind $P_{i,t} \leftarrow t_\text{prev}$  \vspace{0.5mm}\\
      \find $T_{n-1} \leftarrow$ $\text{argmin}_t(C_{n,t})$ \vspace{0.5mm}\\
      \find {\bf for each} $i$ {\bf from} $n - 2$ {\bf to} $1$: \vspace{0.5mm}\\
      \find \ind $T_{i} \leftarrow$ $P_{i+1,T_{i+1}}$ \vspace{0.5mm}\\
    \end{tabular}
  \end{minipage}
  \caption{\emph{Dynamic programming} algorithm for optimizing the
    SymPix grid layout. In summary, the algorithm considers all
    possible solutions, and employ look-up tables of partial solutions
    for bands 0 to $i-1$ when considering band $i$.  The condition
    $P_{i-1,t_\text{prev}} = t_\text{prev}$ ensures that at least two
    bands in a row have the same number of tiles, except (possibly)
    for the first two rows, $T_1 \ne T_0$.}
  \label{code:optimal-grid}
\end{figure}

\subsection{Grid optimization}
\label{sec:algorithm}

We end this section by describing the algorithm used to optimize the
number of of tiles in each band, subject to the constraints defined in
Section \ref{sec:symmetry}. We will in the following only discuss the
Northern hemisphere, as the Southern hemisphere is given directly by
symmetry.

To initialize the algorithm, the user must provide a tile size $k$ and
a total number of rings $\Nrings$, where $\Nrings$ must divisible by
both 2 and $k$. The grid will be able to accurately represent fields
that are band-limited at $\lmax = \Nrings - 1$.  Together, these
parameters specify the angular resolution of the grid, and correspond
in principle to the HEALPix $N_{\textrm{side}}$ parameter. We then
number the bands by $i=0, \dots\, \Nbands-1 \equiv \Nrings/(2k) - 1$, such
that each band consists of $k$ rings. We also define $\alpha_i$ to be
the minimum number of tiles in each band subject to the constraint
that the southmost ring within the band fulfills
Equation~\ref{eq:polar-optimization}.

Deriving the optimal SymPix grid is now equivalent to determining the
number of tiles, $T_i$, for each band. For this optimization process
we adopt the following cost function,
\begin{equation}
  \label{eq:cost}
  c(T_0, \dots, T_{\Nbands - 1}) \equiv \sum_i c_i(T_i) \equiv \sum_i (T_i - \alpha_i)^2,
\end{equation}
which must be minimized subject to
\begin{equation}
  \label{eq:increase-constraint}
  \frac{T_{i+1}}{T_i} \in \left\{\frac{6}{5}, \frac{5}{4}, \frac{4}{3}, 1, 2, 3\right\}.
\end{equation}
Additionally, we initialize the recursion by defining $T_0$ as the
smallest number larger than $\alpha_0$ that is only a product of the
factors $2$, $3$ and $5$, and for computational speed we add the
heuristic (or modification to the cost function) that $T_i <
3\alpha_i$, i.e., that no band should be over-pixelized by more than
three times the Nyquist frequency.

The actual calculation is then a simple exercise in dynamic
programming, as described in any standard text on algorithms
\citep[e.g.][]{introduction-to-algorithms}. Our implementation is
summarized in Figure~\ref{code:optimal-grid}, which has a worst-case
computational complexity of $\mathcal{O}(n\alpha_n) =
\mathcal{O}(\Nrings^2) = \mathcal{O}(\Npix)$, and the same worst-case
memory use. Due to the low computational complexity and the fact that
the optimization only needs to be performed once per grid resolution,
we do not present benchmarks this operation; its computational cost is
negligibly small for our purposes.

\section{Benchmarks and comparisons}

Before considering specific applications, we first characterize the
basic performance of the SymPix grid in terms of computational
efficiency and numerical accuracy.

\subsection{Geometric efficiency}

We start by quantifying the geometric efficiency of our grid, as
characterized by the overall number of grid points and the pixel area
uniformity. For these tests, we consider an example grid with
$\lmax=2000$ and $k=4$, sufficient to discretize a spherical field
with an angular resolution of 15' FWHM. Running the
algorithm summarized in Figure~\ref{code:optimal-grid} with these
input parameters yields a SymPix grid with $5.6\cdot 10^{6}$
grid points.

\begin{figure}
  \begin{center}
    \includegraphics[width=\linewidth]{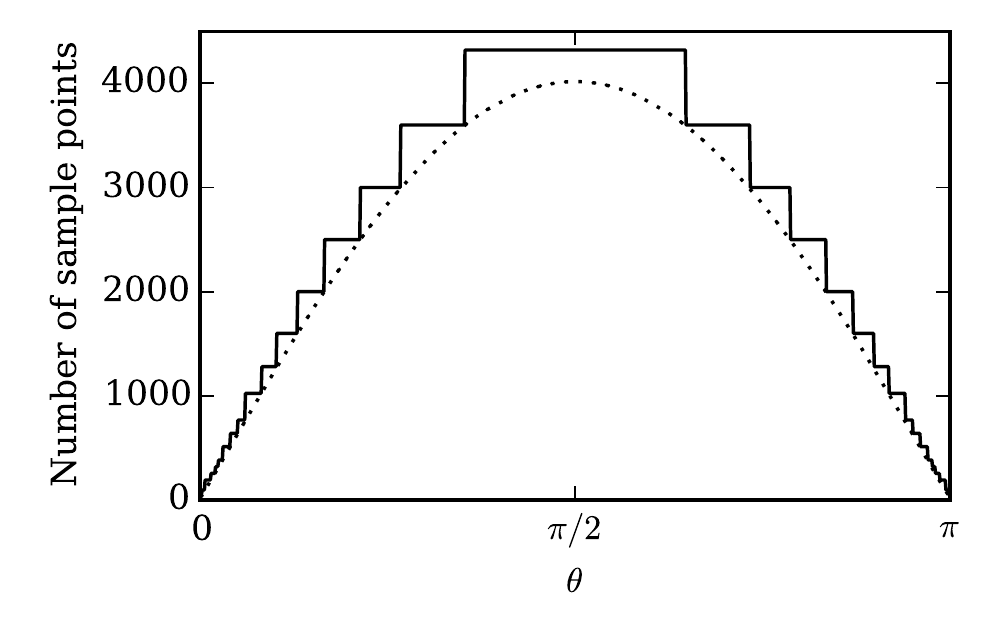}
  \end{center}
  \caption{Number of SymPix grid points per ring as a function of
    latitude (solid line). The dotted line shows $\alpha_i$, i.e., the
    same quantity for the reduced Gauss-Legendre grid
    \citep{reinecke:2013}.}
  \label{fig:ringlens}
\end{figure}

In Figure~\ref{fig:ringlens} we compare the number of SymPix grid
points per ring with the optimal number of points per ring used by the
reduced Gauss-Legendre grid \citep{reinecke:2013}. The ratio between
the solid and dashed lines thus indicates the amount of longitudinal
over-sampling implied by the SymPix grid. Except very close to the
poles, where there are very few points in terms of absolute numbers,
this ratio is never larger than 1.35.

\begin{figure}
  \begin{center}
    \includegraphics[width=\linewidth]{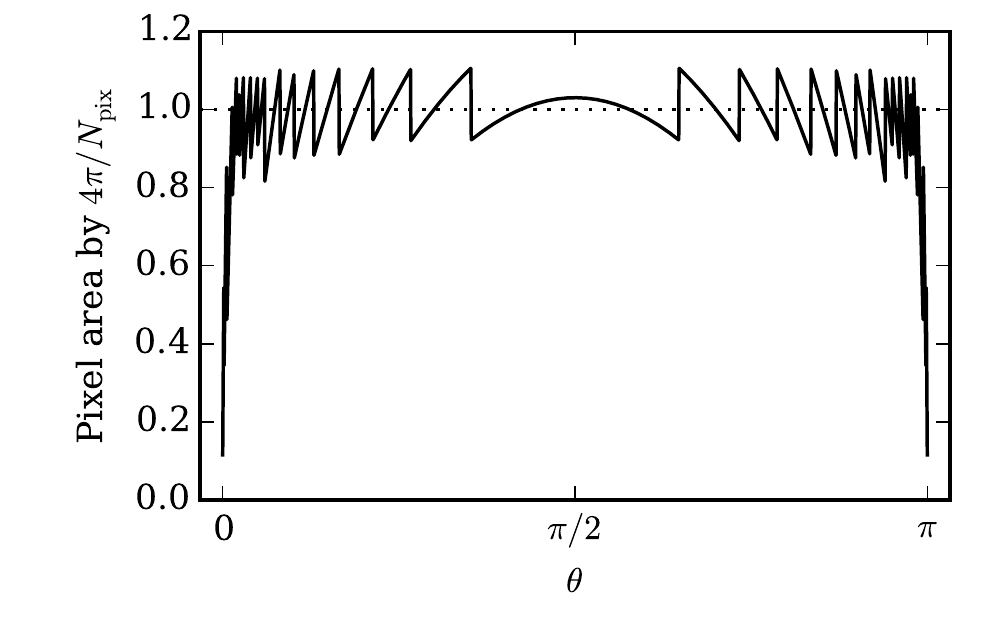}
  \end{center}
  \caption{SymPix pixel area as a function of latitude in units of
    $4\pi/\Npix$ (solid line).  For the HEALPix grid, pixel areas are
    perfectly uniform (dotted line), while significant over-sampling
    occurs close to the poles for the SymPix grid.}
  \label{fig:areas-by-ring}
\end{figure}

A similar illustration is provided in Figure~\ref{fig:areas-by-ring},
where we plot the pixel area as a function of latitude, defining pixel
borders strictly along longitudes and latitudes. The pixel area is
given in units of the pixel area averaged over the full sky, i.e.,
$4\pi/\Npix$, such that a perfectly uniform pixelization, like
HEALPix, corresponds to a constant value of unity. Overall, we see
that the effective pixel areas vary at most by 20\,\% relative to the
average, except near the poles, where the normalized area may be as
low as 0.1.

\begin{figure}
  \begin{center}
    \includegraphics[width=\linewidth]{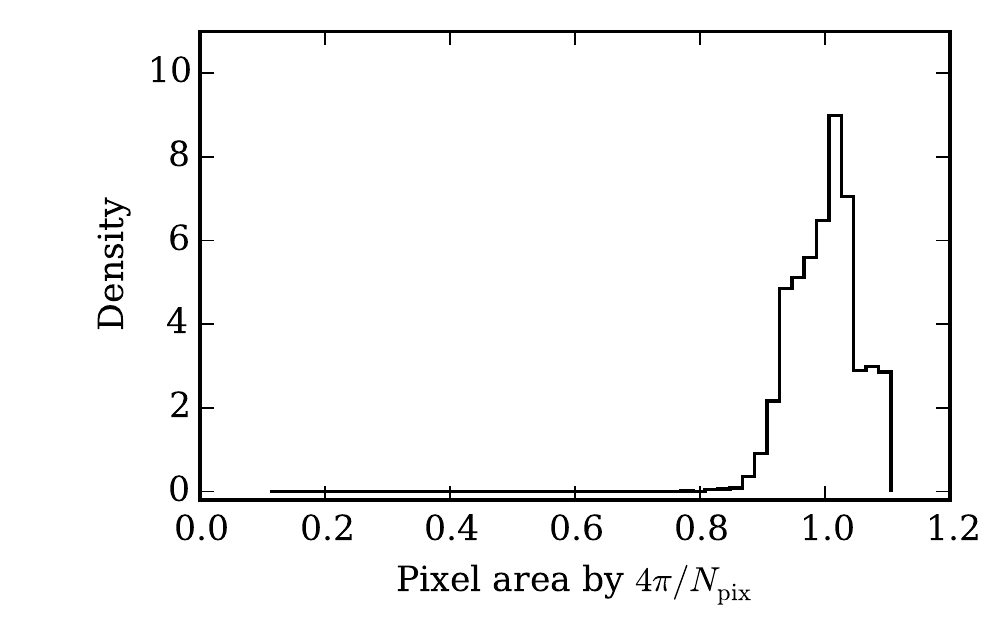}
  \end{center}
  \caption{Histogram of normalized SymPix pixel areas.  The tail
    extending below 0.8 corresponds to polar oversampling, and
    contains about 0.4\% of the total number of pixels for this
    particular grid setup.}
  \label{fig:areas-hist}
\end{figure}

Figure~\ref{fig:areas-hist} shows a histogram of normalized pixel
areas, and we see that the vast majority of grid points have a
normalized area between 0.9 and 1.1. The tail below 0.8 corresponds to
the over-pixelized polar caps, and these contain only 0.4\% of the
total number of grid points for this particular example. Overall, the
SymPix grid implies an over-sampling of about 11\%
compared to the reduced Gauss-Legendre grid, which is acceptable for
our purposes.

\subsection{Accuracy of spherical harmonic quadrature}

\begin{deluxetable*}{clcrcllc}
\tablecaption{\label{tab:comparison}Comparison of different grids
in terms of number of pixels and accuracy of spherical harmonic analysis}
\tablecomments{The HEALPix resolution is kept constant at $\Nside=256$,
while the spherical harmonic bandlimit varies over
$\lmax=\{2.0,2.5,3.0\}\Nside$. The SymPix and Gauss-Legendre
band-limits are identical to the spherical harmonic band-limit. 
}
\tablecolumns{8}
\tablewidth{\linewidth}
\tablehead{
  \colhead{$\lmax$} &
  \colhead{Grid} &
  \colhead{Parameter} &
  \colhead{$\Npix$} &
  \colhead{$\Npix/N_\text{pix}^\text{HEALPix}$} &
  \colhead{Max. error} &
  \colhead{Mean error} &
  \colhead{CPU time for SHT (ms)}
}
\startdata
511 & HEALPix & $\Nside=256$ & 786\,432 & 1.00 & $2.1 \cdot 10^{-2}$ & $2.9 \cdot 10^{-5}$ & 160 \\
 & SymPix & $\lmax=511$ & 390\,656 & 0.50 & $7.8 \cdot 10^{-3}$ & $8.1 \cdot 10^{-7}$ & 67 \\
& Gauss-Legendre & $\lmax=511$ & 524\,288 & 0.67 & $7.5 \cdot 10^{-13}$ & $2.8 \cdot 10^{-14}$ & 66 \\
\\
639 & HEALPix & $\Nside=256$ & 786\,432 & 1.00 & $2.2 \cdot 10^{-1}$ & $1.3 \cdot 10^{-3}$ & 219 \\
 & SymPix & $\lmax=639$ & 591\,232 & 0.75 & $7.2 \cdot 10^{-3}$ & $1.1 \cdot 10^{-6}$ & 118 \\
& Gauss-Legendre & $\lmax=639$ & 819\,200 & 1.04 & $1.2 \cdot 10^{-12}$ & $3.2 \cdot 10^{-14}$ & 118 \\
\\
767 & HEALPix & $\Nside=256$ & 786\,432 & 1.00 & $1.6 \cdot 10^{0}$ & $6.8 \cdot 10^{-2}$ & 287 \\
 & SymPix & $\lmax=767$ & 838\,656 & 1.07 & $4.0 \cdot 10^{-2}$ & $4.8 \cdot 10^{-6}$ & 188 \\
& Gauss-Legendre & $\lmax=767$ & 1\,179\,648 & 1.50 & $1.0 \cdot 10^{-12}$ & $3.8 \cdot 10^{-14}$ & 188
\enddata
\end{deluxetable*}

\begin{figure*}
  \begin{center}
    \includegraphics[width=\linewidth]{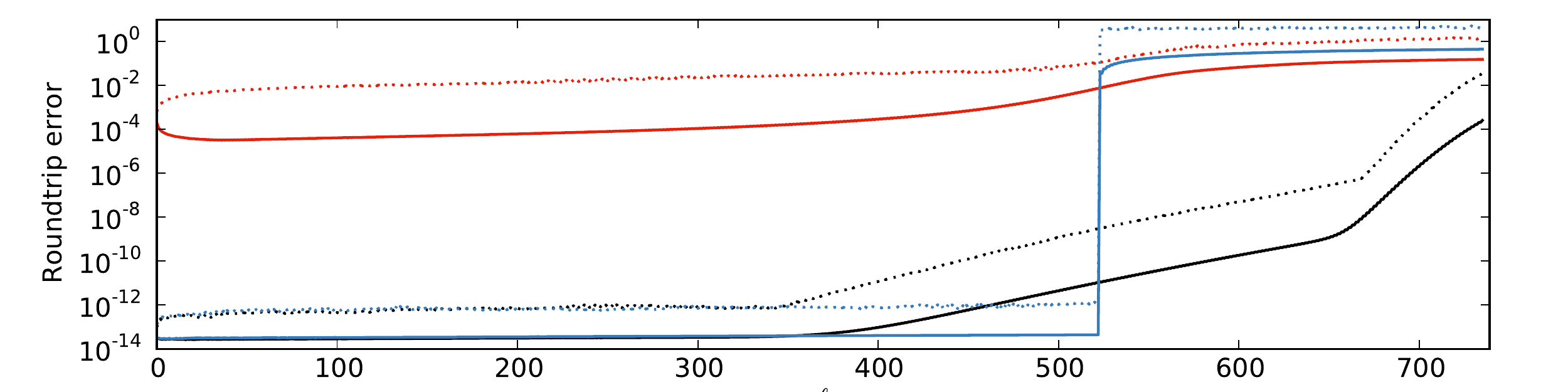}
  \end{center}
  \caption{Spherical harmonic round-trip error as a function of
    multipole, summarized in terms of maximum (dotted lines) and mean
    (solid lines) errors, averaged over both harmonic quantum number
    $m$ and $N_{\textrm{sim}}=100$ simulations. Black lines show
    results for a SymPix grid with $\ell_{\textrm{max}}=735$ and
    tile-size $8$; red lines show results for a HEALPix grid with
    $\Nside=256$ and $\ell_{\textrm{max}}=735$; and blue lines show
    results for a regular Gauss-Legendre grid with
    $\ell_{\textrm{max}}=628$. All grids have roughly the same number
    of grid points, $\Npix \approx 780\,000$.  }
  \label{fig:error-by-l}
\end{figure*}

\begin{figure*}
  \begin{center}
    \includegraphics[width=\linewidth]{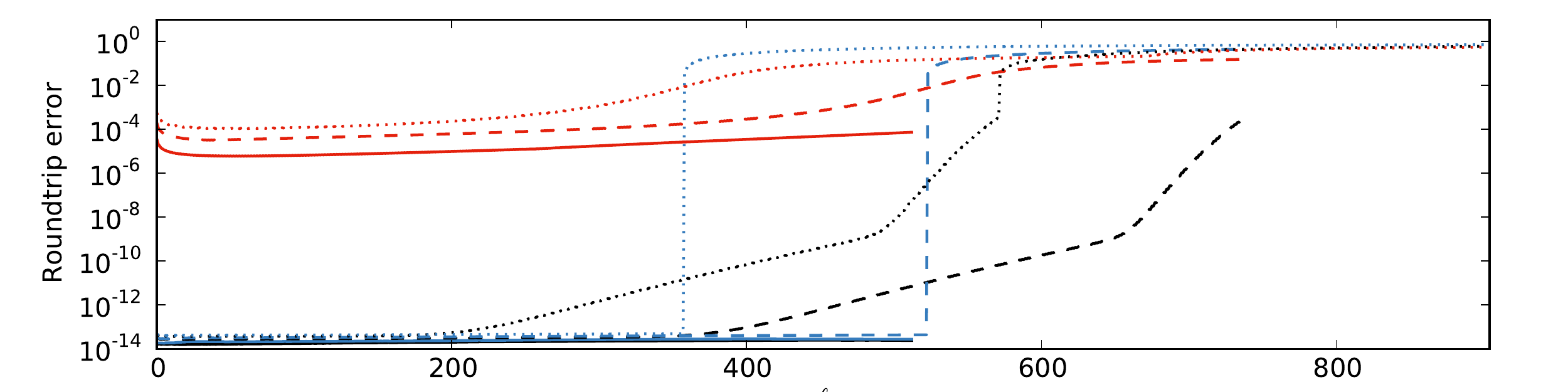}
  \end{center}
  \caption{Error induced by under-sampling (aliasing) as a function of
    multipole in terms of average errors, averaged over both harmonic quantum
    number $m$ and $N_{\textrm{sim}}=100$ simulations. The
    experimental setup is the same as in Figure~\ref{fig:error-by-l},
    but the spherical harmonic bandwidth limit varies between
    $\lmax=512$ (solid), $\lmax=735$ (dashed), and $\lmax=900$
    (dotted).  }
  \label{fig:aliasing}
\end{figure*}

Next, we compare the numerical accuracy of spherical harmonics
transforms as implemented on the SymPix, HEALPix and reduced
Gauss-Legendre grids. This test is carried out through the following
experiment: 
\begin{enumerate}
\item We draw a fiducial signal $\ve{a} = \{a_{\ell m}\}$ in spherical
  harmonic domain, band-limited by some $\lmax$. All spherical
  harmonics coefficients are drawn from the same zero mean and unit
  variance Gaussian distribution, such that no angular scales dominate
  the real-space field.
\item We project this signal onto the respective grid sample points by
  spherical harmonic synthesis.
\item We convert the real-space signal back to harmonic space through
  spherical harmonic analysis, including multipoles up to $\lmax$, to
  recover $\widehat{\a}$.
\item We repeat this procedure $N_{\textrm{sim}}$ times, and summarize
  the results in terms of the resulting round-trip errors,
  $e^{(i)}_\lm\equiv\widehat{a}^{(i)}_\lm - a^{(i)}_\lm$.
\end{enumerate}

Before presenting the results, we note that no fundamental band-limit
and/or resolution parameter $\Nside$ exist for HEALPix for a given
angular resolution. For instance, changing the band-limit $\lmax$ will
add/reduce aliasing for {\em all} scales. A quantitative head-to-head
comparison at a given resolution is therefore difficult, as additional
parameter tuning can affect the results. With this caveat in mind, we
present in Table~\ref{tab:comparison} results for three different
band-limits, $\lmax=\{2.0,2.5,3.0\}\Nside$ with $\Nside=256$, quoting
both the maximum and mean errors as evaluated over all error
coefficients $e^{(i)}_\lm$. Each case includes $N_{\textrm{sim}}=100$
simulations, and the SymPix tile size is fixed at $k=8$.

Starting with the highest bandwidth case, $\lmax=3\Nside$, we first
note that the regular Gauss-Legendre grid is the only one grid that
achieves overall machine precision, with a mean error of
$\mathcal{O}(10^{-14})$ and a maximum error of
$\mathcal{O}(10^{-12})$. For comparison, the corresponding mean and
maximum SymPix errors are $\mathcal{O}(10^{-6})$ and
$\mathcal{O}(10^{-2})$, respectively, while HEALPix achieves
$\mathcal{O}(10^{-1})$ and $\mathcal{O}(1)$ for this high bandwidth
case. Reducing the bandlimit to $\lmax=2\Nside$ improves the latter by
about two orders of magnitude.

However, the statistics listed in Table~\ref{tab:comparison} provide
only a very coarse comparison, because the round-trip errors are
highly scale dependent. In Figure~\ref{fig:error-by-l} we therefore
plot the error as a function of multipole, $\ell$, choosing the SymPix
and HEALPix bandlimits such that the corresponding grids roughly match
a HEALPix $N_{\textrm{side}}=256$ grid in terms of total number of
sample points. For SymPix, this corresponds to $\lmax=735$, and for
the Gauss-Legendre grid it is $\lmax=628$.

Starting with the Gauss-Legendre grid (blue lines), we see that the
error reaches machine precision up to the bandwidth limit; at higher
multipoles no information is carried by the grid. In contrast, the
SymPix grid reaches machine precision up to $\ell \approx 0.5\lmax$,
while the error increases more smoothly at higher multipoles. However,
even though the high-$\ell$ error increase is smooth, it is still
exponential, and the mean and maximum statistics listed in
Table~\ref{tab:comparison} are therefore strongly dominated by the
small-scale errors. Thus, by virtue of deriving its main geometric
grid layout from the Gauss-Legendre grid, we see that the numerical
performance of the SymPix grid is excellent on large and intermediate
angular scales, and the cost of its superior symmetry properties
primarily comes in the form of sub-optimal small-scale residuals. For
comparison, the HEALPix errors are roughly constant at
$\mathcal{O}(10^{-4})$ to $\mathcal{O}(10^{-2})$, and vary only weakly
with angular scale. Note that in all cases the errors can be reduced
by iteration techniques, essentially using least squares minimization
to find the spherical harmonic signal with least power that projects
exactly to the map, and employing the result of spherical harmonic
analysis as a preconditioner. 

The large errors seen for the Gauss-Legendre grid above $\lmax$ is due
to under-sampling or, equivalently, aliasing. In
Figure~\ref{fig:aliasing} we study this effect directly by varying the
spherical harmonics bandwidth limit between $\lmax^{\textrm{SH}}=512$,
735 and 900; note, however, that the actual grid resolution parameters
are kept fixed at the above values, and the higher resolutions
enforced here therefore no longer match the respective grid
properties. Considering first the Gauss-Legendre grid with a SHT
bandlimit of $\lmax=512$, we see, as expected, that the errors reach
machine precision at all scales. However, for the higher bandlimits,
$\lmax=735$ and 900, both of which are higher than the grid resolution
of $\lmax^{\textrm{grid}}=628$, the errors saturate at a multipole
\emph{below} the grid resolution. To be specific, the critical
multipole is $2\lmax^{\textrm{grid}}-\lmax^{\textrm{SH}}$,
corresponding to the well-known aliasing limit from standard Fourier
theory. However, at lower multipoles no aliasing is observed for the
Gauss-Legendre grid, which implies that it is fully robust with
respect to under-sampling, given a known bandlimit.

In comparison, the corresponding HEALPix errors are non-local, in the
sense that increasing the spherical harmonics bandlimit increases the
errors at \emph{all} angular scales: The dotted line ($\lmax=900$)
lies consistently higher than the dashed line ($\lmax=735$), which in
turn lies consistently higher than the solid line ($\lmax=512$). The
HEALPix grid is thus not robust against under-sampling, and it is very
important to choose a grid resolution appropriate for the bandwidth of
the signal under consideration, which in several applications may
imply over-sampling the signal.

The SymPix grid performance lies, as expected, between those of
Gauss-Legendre and HEALPix. On large angular scales, it achieves
numerical precision, while on small scales the aliasing increases
exponentially with multipole, and eventually reaches similar levels as
HEALPix.

\subsection{Computational speed of SHTs}

\begin{figure}
  \begin{center}
    \includegraphics[width=\linewidth]{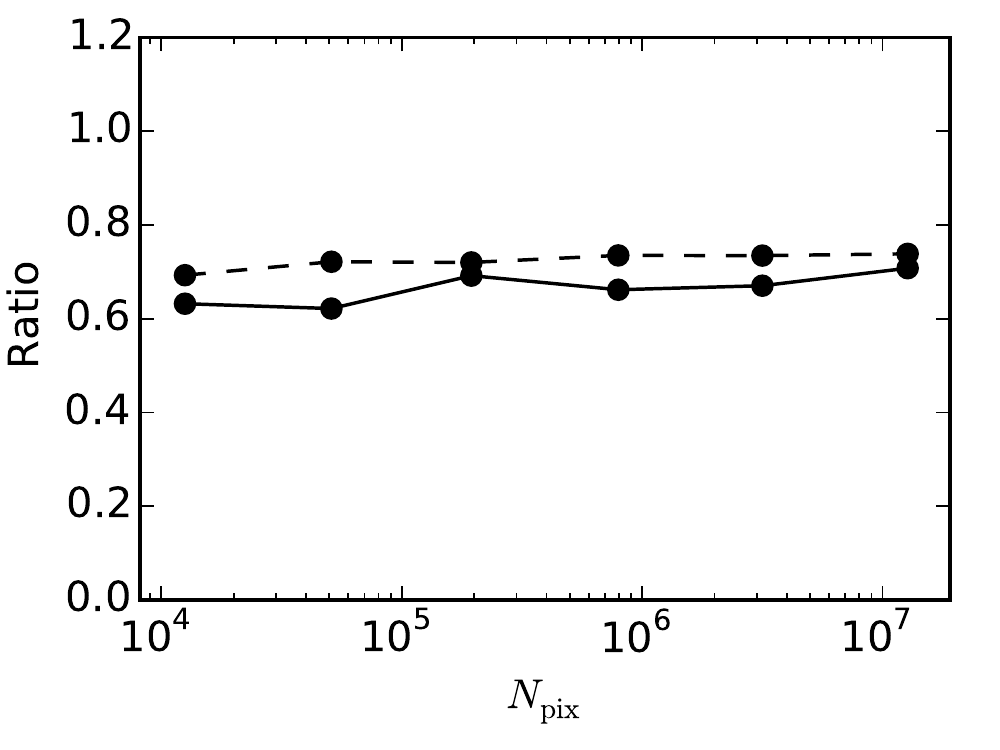}
  \end{center}
  \caption{Comparison between spherical harmonic transforms cost as
    performed with SymPix and HEALPix as a function of $\Npix$, plotted
    in terms of their ratio (black solid line). The dashed line shows
    the ratio between the number of grid point rings.}
  \label{fig:sht-benchmark}
\end{figure}

Before ending this section, we compare the performance of the SymPix,
HEALPix and Gauss-Legendre grids in terms of computational speed. The
rightmost column in Table~\ref{tab:comparison} lists the CPU time for
each of the cases considered above in units of wall-clock
milli-seconds, while Figure~\ref{fig:sht-benchmark} presents a
head-to-head comparison of the SymPix and HEALPix grid performance as
a function of $\Npix$. All benchmarks were performed using {\tt
  libsharp} on a single Intel Core i7 Q840 at 1.87 GHz (SSE2); for
full details including CPU times in absolute numbers, we refer the
interested reader to \citet{reinecke:2013}.

Overall, SymPix perform similarly to the Gauss-Legendre grid, and both
execute about 30\,\% faster than HEALPix. This latter difference may
be explained by the fact that the HEALPix grid points form a zig-zag
pattern in which every other ring is longitudinally shifted by half a
pixel width. This implies a grid point organization that comprise
about 30\,\% more rings than Gauss-Legendre and SymPix grids, which
exhibit more regular longitudinal pixel organizations. This is
relevant, because the computational complexity of SHTs scales as
\begin{align}
  \label{eq:sht-O}
  C_\text{SHT} &= \mathcal{O}(N_\text{ring} \lmax^2) + \mathcal{O}(\Npix \log \frac{\Npix}{N_\text{ring}})
\\\nonumber &= \mathcal{O}(\lmax^3) + \mathcal{O}(\lmax^2 \log \lmax).
\end{align}
The first term represents the cost of computing the associated
Legendre polynomials for each ring, and dominates the second term,
which accounts for evaluating Fast Fourier Transforms (FFTs) along
each ring. Thus, the number of grid points per ring is not critical
for the overall speed of SHTs, while the total number of rings is.

In addition, SymPix grids have by construction rings with pixel
numbers that are only products of $2$, $3$ and/or $5$, which ensures
efficient Fast Fourier Transforms (FFTs). In contrast, many HEALPix
rings have pixel numbers that includes large primes, and therefore the
Bluestein algorithm must be employed for these. This effect is more
important for lower resolution grids, for which the cost of FFTs is
relatively higher.

\section{Applications}

We now turn our attention to practical applications, and in particular
to the construction of efficient preconditioners. Before doing that,
however, we consider a simpler application, namely real-space
convolution, in order to build up intuition regarding the relevant
operations. We emphasize that the purpose of this preliminary
discussion is not to provide a real-world alternative to spherical
harmonic transforms, or the methods presented by \cite{elsner} and
\cite{sutter} for such convolutions, but simply to quantify the
computational efficiency of the SymPix grid on a simple and intuitive
application.

\subsection{Spherical convolution}
\label{sec:convolution}

The convolution of a spherical image $f$ with a kernel $b$
is given by the spherical surface integral
\begin{equation}
  \label{eq:convolution}
  g(\hat{n}) = \int_{4\pi} b(\hat{n},\hat{m}) f(\hat{m}) d\Omega_{\hat{m}}.
\end{equation}
In our case we assume an azimuthally symmetric kernel, and
$b(\hat{n},\hat{m})$ therefore depends only on the distance between
$\hat{n}$ and $\hat{m}$, such that
\begin{equation}
  \label{eq:convolution-rewritten}
  g(\hat{n}) = \int_{4\pi} b(\hat{n} \cdot \hat{m}) f(\hat{m}) d\Omega_{\hat{m}}.
\end{equation}
This integral is most commonly performed in spherical harmonic domain,
turning full-sky convolution into coefficient-wise multiplication with
a corresponding transfer function, $b_\ell$, which is given by the
Legendre transform of $b(\hat{n} \cdot \hat{m})$. These computations
are dominated by the spherical harmonic transforms, and therefore have
a computational scaling of $\mathcal{O}(\Npix^{3/2}) =
\mathcal{O}(\lmax^3)$.

If $b$ is spatially narrow compared to the required pixelization, as
is usually the case, one could instead consider the pixel-domain
convolution by evaluating
\begin{equation}
  \label{eq:convolution-pixel}
  g(\hat{n}_i) = \sum_{j=1}^{\Npix} b(\hat{n}_i \cdot \hat{n}_j) f(\hat{n}_j),
\end{equation}
where the convolution kernel reads
\begin{align}
\label{eq:addition-theorem}
b(x) = \sum_{\ell=0}^{\lmax} \frac{2\ell + 1}{4\pi} b_\ell P_\ell(x).
\end{align}
One would then make the approximation that $b(\hat{n}_i \cdot
\hat{n}_j) = 0$ whenever sample points $i$ and $j$ are more than $k$
sample point distances apart, as discussed in
Section~\ref{sec:introduction}. 

For HEALPix, almost all sample point distances are different, and $b$
must therefore be evaluated $\mathcal{O}(\Npix\,k^2)$ times. The
computational complexity of pixel-domain convolution on the HEALPix
grid therefore scales as $\mathcal{O}(\Npix\,k^2 \,\lmax) =
\mathcal{O}(k^2 \,\lmax^3)$, which is clearly inferior to the harmonic
approach both in terms of speed and accuracy. With SymPix, however,
the large number of symmetries allows us to reduce the computational
complexity to $\mathcal{O}(k^2\,\Npix + \sqrt{\Npix}\,\lmax) =
\mathcal{O}(k^2 \,\Npix)$: One simply needs to choose a tile size $k$
such that only sample point pairs within a tile and between
neighbouring tiles must be considered. Then for, each band of $k$
rings, $b(\hat{n} \cdot \hat{m})$ only needs to be evaluated for the
first few tiles of the band, as other distances within the same band
will be identical within the remainder of the band.

The speed-up for evaluating all necessary $b(\hat{n} \cdot
\hat{m})$, when approximating $b(\hat{n} \cdot \hat{m}) = 0$ whenever
$\hat{n}$ and $\hat{m}$ are not in neighbouring tiles, are given in
Table \ref{tab:sample_benchmark}.  In addition to scaling better than
the $\mathcal{O}(\Npix^{3/2})$ spherical harmonic transforms, this
approach should also be easier to parallelize and implement
efficiently on a GPU.

\begin{deluxetable}{ccc}
\tablecaption{\label{tab:sample_benchmark}CPU time and theoretical speed-up for 
  evaluating $b(\hat{n} \cdot \hat{m})$}
\tablecomments{%
We have approximated $b(\hat{n} \cdot \hat{m}) = 0$ whenever $\hat{n}$ and $\hat{m}$
are not in neighbouring tiles.
The third column shows the number of non-zero $b(\hat{n} \cdot \hat{m})$,
which scales as $O(k^2 \Npix)$, divided by the number of elements we
had to compute when making use of the SymPix symmetries, which scales as $O(k^2 \sqrt{\Npix})$.
In this example we have chosen $k=8$.
}
\tablecolumns{3}
\tablewidth{0.9\linewidth}
\tablehead{
  &
  \colhead{CPU time}
  &
  \colhead{Speed-up}
  \\
  \colhead{$\lmax$}&\colhead{[sec]} &\colhead{[factor]}
}
\startdata
3000&    9.8\phn    & 732 \\
1500&    3.6\phn    & 335 \\
\phn750& 1.4\phn    & 149 \\
\phn375& 0.74       & \phn70 \\
\phn188& 0.50       & \phn26 \\
\phn100& 0.31       & \phn14
\enddata
\end{deluxetable}

Note that yet another method for spherical convolution with a
symmetric kernel has been implemented in the ARKCoS code
\citep{elsner,sutter}, with a computational scaling of $\mathcal{O}(k
\,\lmax^2 \,\text{log}\,\lmax) = \mathcal{O}(k \,\Npix \,\log
\,\Npix)$. Whether a SymPix-based convolution would improve relative
to their work for relevant resolution parameters and accuracy
requirements remains to be explored.

\subsection{Preconditioner construction for linear systems}
\label{sec:solvers}

Finally, we are in the position to discuss the application of the
SymPix grid to our main usecase, namely for solving linear systems
involving rotationally invariant operators in pixel domain, either
through multi-grid methods or to construct efficient preconditioners.
The simplest example of such a system is
\begin{equation}
  \label{eq:simple}
  \Y \B \Y^T \ve{x} = \ve{b},
\end{equation}
where $\Y$, as usual, is the matrix corresponding to spherical
harmonic synthesis and $\B$ is a diagonal matrix in spherical harmonic
domain, $B_{\ell m,\ell' m'} = b_\ell \delta_{\ell,\ell'}
\delta_{mm'}$. The product $\Y \B \Y^T$ is a pixel domain operator
with strong spatial couplings within the correlation length implied by
$b$. Of course, this particular system could have been trivially
solved by converting to spherical harmonic domain, which would
diagonalize the coefficient matrix. However, if there are more terms
in the operator, this is no longer possible, and iterative solvers
like Conjugate Gradients or multi-level algorithms are needed. In
these cases SymPix is useful to construct preconditioners or
smoothers.

Our own main interest lies in drawing constrained Gaussian
realizations of the CMB sky by using a multi-level solver
\citep{multigrid}.  This may performed by solving the following linear
system \citep{jewell:2004,wandelt:2004,eriksen:2004},
\begin{equation}
  \label{eq:cr-system-level}
 \Y_1(\D + \B \Yobs \N^{-1} \Yobs^T \B )\Y_1^T \ve{x} = \ve{r},
\end{equation}
where $\ve{D}$ and $\ve{B}$ are diagonal matrices in spherical
harmonic domain, characterized by transfer functions $d_\ell$ and
$b_\ell$, $\N^{-1}$ is a diagonal (inverse noise covariance) matrix in
pixel domain, pixelized on some external grid $\theta_i$, and $\ve{r}$
is a stochastic term that depends on the data set in question.

Two different spherical grids are involved in system. First, the
outermost spherical harmonics transform, $\Y_1$, denotes synthesis to
a grid of our own choosing. We will use a SymPix grid of resolution
$\lmax$ for this operator in the following.  The inner transform,
$\Yobs$, is determined by some external experiment, and is thus not
flexible. Here we will assume that this operator is defined on a
full-sky HEALPix grid of $\Nside=2048$, typical for the CMB maps
published by the \emph{Planck} experiment \citep{planck2014-a01}.

Of course, from the viewpoint of the overall linear system, the
details of any individual operator is irrelevant, and the only crucial
point is that the combined operator remains the same. In order to
speed up the calculations through use of symmetries, we
therefore substitute the inner-most HEALPix based noise covariance
matrix product with a corresponding SymPix based product,
\begin{equation}
  \label{eq:substitution}
  \Yobs \N^{-1} \Yobs^T = \Y_2 \N_{2}^{-1} \Y_2^T
\end{equation}
where $\Y_2$ denotes an auxiliary SymPix grid; note that this does not
need to be the same as $\Y_1$, but its resolution can be adjusted to
trade numerical precision for computational speed. As shown by
\citet{multigrid}, Equation~\ref{eq:substitution} holds true if $\N_2$
is constructed from
\begin{equation}
  \label{eq:degrade}
  \ve{\theta}_2 = \W_2 \Y_2 \Yobs^T \ve{\theta},
\end{equation}
in the same way as $\N$ is constructed from $\theta$. In this latter
expression, $\W_2$ is a diagonal matrix containing the quadrature
weights used in the spherical harmonic analysis of the target grid,
while $\Yobs^T$ lacks the ring weights one normally uses in spherical
harmonic analysis. Note that this operation is in fact the opposite
procedure compared to naive resampling, which would be written $\Y_2
\Yobs^T \W_\text{obs}$ in our notation. For full details, we refer the
interested reader to \cite{multigrid}.

\begin{deluxetable}{lccc}
\tablecaption{\label{tab:precond_benchmark}CPU time for constructing
  preconditioner}
\tablecomments{The top section lists the CPU time for preconditioner
  calculations that depend only on data geometry 
  (mask, beam, noise characterization), while the bottom section lists
  the corresponding CPU time for calculations that depend on
  $d_{\ell}$, which in CMB applications typically corresponds to an
  angular power spectrum, $C_{\ell}$. The second column is copied
  directly from \citet{multigrid}, and shows results using HEALPix for
  all calculations. The third row shows similar results using SymPix,
  while the fourth column
  shows the ratio between the two.}
\tablecolumns{4}
\tablewidth{0.95\linewidth}
\tablehead{
  \colhead{} &
  \colhead{HEALPix} &
  \colhead{SymPix} &
  \colhead{}\\
  \colhead{$\lmax$} &
  \colhead{(CPU min)} &
  \colhead{(CPU min)} &
  \colhead{Speed-up}
}
\startdata
\cutinhead{Evalution of $\widehat{\B}^T \N^{-1} \widehat{\B}$}
3000  & 727 & 5.4\phn\phn    & \phantom{\,}\phn\phn130 \\
1500  & 509 & 1.4\phn\phn    & \phantom{\,}\phn\phn360 \\
\phn750   & 340 & 0.37\phn   & \phantom{\,}\phn\phn920 \\
\phn375   & 230 & 0.11\phn   & \phn2\,100 \\
\phn188   & 452 & 0.035  & 13\,000\\\vspace*{1mm}
\phn100   & 363 & 0.027  & 13\,000\\\vspace*{-1mm}
Sum     & 2\,621\phn & 7.3\phn\phn  & \phantom{\,}\phn\phn360\\
\cutinhead{Evalution of $\widehat{\D}$}
3000  & \phn\phn\phn85\phantom{.}\phn\phn\phn   & 3.3\phn   & \phantom{\,}\phn\phn\phn26 \\
1500  & \phn\phn\phn15\phantom{.}\phn\phn\phn   & 0.83  & \phantom{\,}\phn\phn\phn18 \\
\phn750   & \phn\phn\phn2.4\phn  & 0.22  & \phantom{\,}\phn\phn\phn11 \\
\phn375   & \phn\phn\phn0.36 & 0.07 & \phantom{\,}\phn\phn\phn\phn5 \\
\phn188   & \phn\phn\phn0.05 & 0.02  & \phantom{\,}\phn\phn\phn\phn3 \\\vspace*{1mm}
\phn100   & \phn\phn\phn0.01 & 0.01  & \phantom{\,}\phn\phn\phn\phn1 \\\vspace*{-1mm}
Sum     &  \phn\phn\phn103\phantom{.}\phn\phn\phn\phn   & 4.5\phn  & \phantom{\,}\phn\phn\phn23
\enddata
\end{deluxetable}

The precision of Equation~\ref{eq:substitution} depends on the
relative bandlimit of $\Y_1$, $\Y_2$ and $\Yobs$. For instance,
choosing $\lmax$ for $\Y_2$ and $\Yobs$ to be twice that of $\Y_1$
yields a numerical precision of $\mathcal{O}(10^{-10})$. Increasing
these to four times that of $\Y_1$ results in an accuracy of
$\mathcal{O}(10^{-14})$, whereas reducing it to only one, such that
$\Y_1 = \Y_2$, gives an accuracy of $\mathcal{O}(10^{-2})$. Even the
latter may be acceptable for preconditioning purposes.

In order to derive an approximation to the full coefficient matrix
defined by Equation~\ref{eq:cr-system-level}, we first re-write the
system as
\begin{equation}
  \label{eq:system-rewritten}
  \widehat{\D} + \widehat{\B}^T \N_2^{-1} \widehat{\B} \ve{x} = \ve{r},
\end{equation}
where
\begin{equation}
  \label{eq:Dhat}
  \widehat{\D} = \Y_1 \D \Y_1^T \quad \text{and} \quad
  \widehat{\B} = \Y_2 \B \Y_1^T.
\end{equation}
We now introduce the approximation that $\widehat{D}_{ij} = 0$ and
$\widehat{B}_{ij} = 0$ whenever two sample points $i$ and $j$ are not
in the same or neighbouring tiles, as per the SymPix organization.
The non-zero elements (i.e., the ``local'' part) of $\widehat{\D}$ and
$\widehat{\B}$ are evaluated by Equation~\ref{eq:addition-theorem}, at
a cost of $\mathcal{O}(\lmax)$ operations per matrix element. However,
as discussed in Section~\ref{sec:convolution}, evaluating all required
elements for a SymPix grid scales as $\mathcal{O}(k^2\,
\sqrt{\Npix})$, as opposed to $\mathcal{O}(k^2 \,\Npix)$ for less
symmetric grids.

These calculations constitute essential components of the
pre-computation step of the multi-grid solver presented by
\citet{multigrid}. In that paper, all evaluations were performed with
the HEALPix grid, with a computational scaling of $\mathcal{O}(\lmax
k^2 \Npix)$ as discussed above. Their Table~2 summarizes the resulting
computational costs in units of CPU minutes. Here we repeat those
calculations adopting exactly the same overall parameters,
facilitating a one-to-one comparison, but we employ SymPix for
intermediate calculations instead of HEALPix. The results are
summarized in Table~\ref{tab:precond_benchmark}, in which the second
column is copied directly from \citet{multigrid}, and the third column
shows the new SymPix results. The fourth column shows the ratio
between the two.

Clearly, the net gains achieved by the SymPix grid varies with
resolution. For the high resolution levels the speed-up is driven by
symmetries drastically reducing the time taken to evaluate
$\widehat{\B}$. The theoretical speed-up of 732 times for evaluating
$\widehat{\B}$ at $\lmax=3000$, found in Table
\ref{tab:sample_benchmark}, is reduced to 130 and 26 for
$\widehat{\B}^T \N^{-1} \widehat{\B}$ and $\widehat{\D}$,
respectively. This is due to work that was previously unimportant now
dominating the computation.  At lower resolutions the speed-up is
almost entirely due to being able to use the operator resampling given
in Equation \eqref{eq:substitution}. This degradation procedure is not
possible when using the HEALPix grid, and so our previous code had to
use a resolution of $\Nside=2048$ along columns and level resolution
along rows.

Overall, the SymPix grid reduces what used to be over-night jobs to
essentially interactive tasks.

\section{Conclusion}

We have presented SymPix, a novel spherical grid for efficient
sampling of rotationally invariant operators. This grid derives many
of its properties from the Gauss-Legendre grid, ensuring overall
excellent spherical harmonics transform performance. The main
difference between the two grids is that SymPix sacrifices proper
Nyquist sampling in the longitudinal direction in order to increase
pixel symmetries, such that all grid pair distances repeat perfectly
along constant-latitude rings. This decreases the computational
scaling of evaluating rotationally invariant operators from
$\mathcal{O}(\Npix)$ to $\mathcal{O}(\sqrt{\Npix})$.

The intended primary application of the SymPix grid is efficient
construction of preconditioners (or smoothers) for iterative linear
solvers. In this paper we considered the specific example of drawing
constrained Gaussian realizations using a multi-grid solver, which is
an important problem in current CMB analysis. Comparing with previous
state-of-the-art results based on the HEALPix grid \citep{multigrid},
we achieve average speed-ups of 360 and 23 for the two most important
pre-computation steps when using SymPix for internal calculations.

However, we emphasize that SymPix is a special-purpose grid designed
for precisely such tasks; it is not intended to provide a general
purpose spherical pixelization that is suitable for, say, map
making. HEALPix is clearly preferred for such purposes due to its
uniform pixel areas, regular pixel window and hierarchical pixel
structure. Likewise, if machine precision spherical harmonics
transforms are required, the Gauss-Legendre grid is the obvious
choice. However, for those particular applications that can benefit
from efficient pixel space sampling of linear operators, such as ours,
SymPix holds a clear edge over existing alternatives.

\begin{acknowledgements}
  DSS and HKE are supported by European Research Council grant
  StG2010-257080.
\end{acknowledgements}

\appendix

\section{Code}

The SymPix code has been developed as part of the Commmander project,
and does not yet have its own library. For the benefit of the reader we have
however copied the source files relevant to this paper
to its own repository at {\tt http://github.com/dagss/sympix}. Please consult
the accompanying README file for further details. This repository will be
updated if the code does eventually develop into a stand-alone package.

The SHTs are all done using {\tt libsharp} \citep{reinecke:2013}, at the
time of writing available at {\tt http://sourceforge.net/projects/libsharp/}.
We then construct the grid geometry in our Python code and feed it to {\tt libsharp}.
In the future we may port our Python code to C and make it available directly in
{\tt libsharp}.

\end{document}